**Title:** Point-surface fusion of station measurements and satellite observations for mapping PM$_{2.5}$ distribution in China: methods and assessment

**Authors:** Tongwen Li [a], Huanfeng Shen [a,b,c,*], Chao Zeng [d], Qiangqiang Yuan [e], Liangpei Zhang [b,f]

**Affiliations:**

[a] School of Resource and Environmental Sciences, Wuhan University, Wuhan, Hubei, 430079, China.

[b] The Collaborative Innovation Center for Geospatial Technology, Wuhan, Hubei, 430079, China.

[c] The Key Laboratory of Geographic Information System, Ministry of Education, Wuhan University, Wuhan, Hubei, 430079, China.

[d] The State Key Laboratory of Hydroscience and Engineering, Department of Hydraulic Engineering, Tsinghua University, Beijing, 100084, China.

[e] School of Geodesy and Geomatics, Wuhan University, Wuhan, Hubei, 430079, China.

[f] The State Key Laboratory of Information Engineering in Surveying, Mapping and Remote Sensing, Wuhan University, Wuhan, Hubei, 430079, China.

**Email address:**

litw@whu.edu.cn (Tongwen Li), shenhf@whu.edu.cn (Huanfeng Shen), zengchaozc@hotmail.com (Chao Zeng), yqiang86@gmail.com (Qiangqiang Yuan), zlp62@whu.edu.cn (Liangpei Zhang)

[*] **Corresponding author:**

Huanfeng Shen (shenhf@whu.edu.cn). Phone: +86-027-68778375

**ABSTRACT**

Fine particulate matter (PM$_{2.5}$, particulate matters with aerodynamic diameter less than 2.5 $\mu m$) is associated with adverse human health effects, and China is currently suffering from serious PM$_{2.5}$ pollution. To obtain spatially continuous ground-level PM$_{2.5}$ concentrations, several models established by point-surface fusion of ground station and satellite observations have been developed. However, how well do these models perform at national scale in China? Is there space to improve the estimation accuracy of PM$_{2.5}$ concentration? The contribution of this study is threefold. Firstly, taking advantage of the newly established national monitoring network, we develop a national-scale generalized regression neural network (GRNN) model to estimate PM$_{2.5}$ concentrations. Secondly, different assessment experiments are undertaken in time and space, to comprehensively evaluate and compare the performance of the widely used models. Finally, to map the yearly and seasonal mean distribution of PM$_{2.5}$ concentrations in China, a pixel-based merging strategy is proposed. The results indicate that the conventional models (linear regression, multiple linear regression, and semi-empirical model) do not perform well at national scale, with cross-validation R values of 0.488~0.552 and RMSEs of 30.80~31.51 $\mu g/m^3$, respectively. In contrast, the more advanced models (geographically weighted regression, back-propagation neural network, and GRNN) have great advantages in PM$_{2.5}$ estimation, with R values ranging from 0.610 to 0.816 and RMSEs from 20.93 to 28.68 $\mu g/m^3$, respectively. In particular, the proposed GRNN model obtains the best performance. Furthermore, the mapped PM$_{2.5}$ distribution retrieved from 3-km MODIS aerosol optical depth (AOD) products, agrees quite well with the station measurements. The results also show that our study has the capacity to provide reasonable information for the global monitoring of PM$_{2.5}$ pollution in China.

**Keywords:** Satellite remote sensing; Point-surface fusion; AOD; PM$_{2.5}$; GRNN; Assessment

# 1. Introduction

Fine particulate matter (PM$_{2.5}$, particulate matters with aerodynamic diameter less than 2.5 $\mu m$) can carry toxic and harmful substances and travel across countries and geographic boundaries (Engel-Cox et al., 2013). Many epidemiological studies have shown that long-term exposure to PM$_{2.5}$ is associated with adverse health effects (Bartell et al., 2013; Sacks et al., 2011). With the rapid economic development, China is suffering from serious air pollution, and PM$_{2.5}$ has gradually become the primary pollutant, which has attracted widespread social concern (Peng et al., 2016; Yuan et al., 2012). Consequently, PM$_{2.5}$ has been incorporated into the new air quality standard of the Chinese government (GB 3095-2012). Since January 2013, hourly PM$_{2.5}$ concentrations have been disclosed to the public through the Chinese National Environmental Monitoring Center website (http://www.cnemc.cn). By the end of 2014, about 1500 monitoring sites had been established to report the overall air quality in China.

Despite the high precision and stability, there are still some limitations to spatiotemporal analysis due to the sparse and uneven distribution of the ground stations. Unlike the ground-level measurements, satellite-based observation has the capacity to provide wide-coverage data. Using both the ground station measurements and co-located satellite observations, the relationship between the various observed variables can be established. Based on this relationship and its variation rule in space, the spatially continuous data can be reconstructed. This method, which can generate data from point scale to surface scale, is known as "point-surface fusion". The point-surface fusion of station-level PM$_{2.5}$ measurements and satellite-based aerosol optical depth (AOD, also called aerosol optical thickness) can obtain spatially continuous $PM_{2.5}$ data, and has the potential to compensate for the spatiotemporal limitation. Several widely used models, established by point-surface fusion, have been developed to describe the relationship between PM and AOD (AOD-PM relationship) (Beloconi et al., 2016; Chu et al., 2003; Gupta et al., 2006; Hoff and Christopher, 2009; Kloog et al., 2014; Li et al., 2011; Li et al., 2005; Liu et al., 2007; Martin, 2008).

According to a previous study (Lin et al., 2015), the existing models, which were developed to retrieve ground-level $PM_{2.5}$ concentrations using satellite observations, can be classified into two main categories: simulation-based models and observation-based models. Simulation-based models (Geng et al., 2015; Liu et al., 2004; van Donkelaar et al., 2010) consider the effects of both meteorology and aerosol properties, which are simulated with global or regional chemical transport models. They would be most suitable for predicting $PM_{2.5}$ concentration if we have comprehensive datasets (especially emission inventories) and a good understanding of the $PM_{2.5}$ formation and removal processes. Given the complexity of the problem, an observation-based model is a good compromise (Gupta and Christopher, 2009a). Observation-based models rely on the statistical relationship between AOD and in-situ $PM_{2.5}$ measurements, and are much easier to implement, but with an almost equivalent accuracy of $PM_{2.5}$ estimation. Hence, the observation-based models, established by point-surface fusion, have been extensively discussed and studied. Using a simple linear regression model between AOD and $PM_{2.5}$, early studies obtained some reasonable results (Chu et al., 2003; Guo et al., 2009; Li et al., 2005; Wang and Christopher, 2003; Wang et al., 2010). However, the relationship tends to be influenced by region and time due to the effects of variations in emissions and meteorological conditions. Through incorporating more meteorological parameters (e.g., relative humidity, temperature, wind speed), a multiple linear regression model may better represent the AOD-PM$_{2.5}$ relationship (Benas et al., 2013; Gupta and Christopher, 2009b). Unlike the empirical models, semi-empirical models take the related physical understanding into account (Emili et al., 2010; Liu et al., 2005; Tian and Chen, 2010; You et al., 2015b), and attempt to introduce physical prior knowledge to solve the problem. More recently, allowing for the spatial heterogeneity of the AOD-PM$_{2.5}$ relationship, a more advanced statistical model called geographically weighted regression (GWR) has been developed to estimate PM$_{2.5}$ concentration (Hu et al., 2013; Ma et al., 2014; Song et al., 2014; You et al., 2015b). This model predicts PM$_{2.5}$ concentration using a local regression approach instead of globally constant regression parameters. In addition, as one of the intelligent algorithms, artificial neural networks (ANNs) have the potential to better represent the complex nonlinear relationship. Hence, ANNs have been introduced into the estimation of PM$_{2.5}$ concentration (Gupta and Christopher, 2009a; Wu et al., 2012; Yao and Lu, 2014), which has been gradually considered to be a multi-variable and nonlinear problem. Furthermore, some more complex mixed effects models and generalized additive mixed models (GAM) have been developed (Kloog et al., 2011; Liu et al., 2009; Ma et al., 2016). On the other hand, considering the effect of the main aerosol characteristics, an observation-based method was developed by establishing a multi-parameter remote sensing formula of PM$_{2.5}$ concentration (Li et al., 2016; Lin et al., 2015; Zhang and Li, 2015). All these observation-based models have been widely used, and have played an irreplaceable role in satellite-based estimation of PM$_{2.5}$ concentration.

China is now facing a serious PM$_{2.5}$ pollution problem (Peng et al., 2016; Zhang and Cao,

2015). Due to the wide geographical range and complex terrain, mapping the distribution of $PM_{2.5}$ concentration in China is faced with lots of challenges. To date, many researchers have made attempts to study the AOD-PM relationship in China (Guo et al., 2009; Li et al., 2011; Li et al., 2005; Lin et al., 2015; Ma et al., 2014; Ma et al., 2016; Song et al., 2014; Wang et al., 2010; Wu et al., 2012). Due to the unavailability of sufficient $PM_{2.5}$ measurements in China before 2013, most studies used a limited number of ground-level $PM_{2.5}$ measurements at a regional scale. However, the regional estimation and analysis of $PM_{2.5}$ concentration cannot provide sufficient information for the macroscopical monitoring of the whole of China. With the newly available national $PM_{2.5}$ measurements since January 2013, a few attempts have been made to estimate $PM_{2.5}$ concentration at national scale. However, their methods and data used to establish the AOD-$PM_{2.5}$ relationship differ greatly from one another. Additionally, the validation schemes of models have many differences; for instance, some schemes undertook validation based on yearly/monthly average, and some on a daily basis. Thus, the intercomparison of national-scale model performance is not possible. Furthermore, previous studies (Hoff and Christopher, 2009) have suggested that the models may perform differently in different regions. As a result, the performance of models developed in regional studies needs to be evaluated and compared at national scale. On the other side, although the widely used models have achieved reasonable results under certain conditions, the estimation accuracy of $PM_{2.5}$ concentration still has room for improvement (De Leeuw et al., 2006; Song et al., 2014).

In this paper, one of the main objectives is to introduce an advanced model (generalized regression neural network, GRNN), which can better represent the AOD-$PM_{2.5}$ relationship, into the prediction of PM$_{2.5}$ distribution in China. Another main objective is to comprehensively evaluate and analyze the performance of the widely used mainstream models at national scale. Finally, a direct average only reflects the level of PM$_{2.5}$ pollution on those days with valid AOD data, so a pixel-based merging strategy is proposed to map the yearly and seasonal mean distribution of PM$_{2.5}$ concentrations.

**2. Data and measurements**

To estimate PM$_{2.5}$ concentration in China, the retrieval models were established using multi-source data. Some information of the various data sources is shown in Table 1, and further details are provided in Sections 2.1 to 2.3.

Table 1. Some information of various data used in statistic models to estimate PM$_{2.5}$ in China

| Data | Frequency | Source |
| --- | --- | --- |
| PM$_{2.5}$ | Daily average | About 1500 air quality monitoring sites (in 2014) in China mainland |
| AOD | Daily | MODIS Terra and Aqua satellite |
| Meteorological data | Daily | MERRA-2 reanalysis data |

*2.1. Ground-level PM$_{2.5}$ measurements*

Daily average PM$_{2.5}$ concentration data from February 2013 to December 2014 were obtained from the Chinese National Environmental Monitoring Center (CNEMC) website (http://www.cnemc.cn). With a spatial coverage of the whole of China (excluding Hong Kong, Macao, and Taiwan), the monitoring network is being continuously updated, and the number of monitoring sites has increased from ~500 in early 2013 to ~1500 by the end of 2014. According to the Chinese National Ambient Air Quality Standard (CNAAQS, GB3905-2012), the ground-level PM$_{2.5}$ concentration should be measured by the tapered element oscillating microbalance method (TEOM) or with beta attenuation monitors (BAM or beta-gauge), with an uncertainty of 0.75% for the hourly record (Engel-Cox et al., 2013; Lin et al., 2015). Fig. 1 shows the spatial distribution of PM$_{2.5}$ monitoring sites in China.

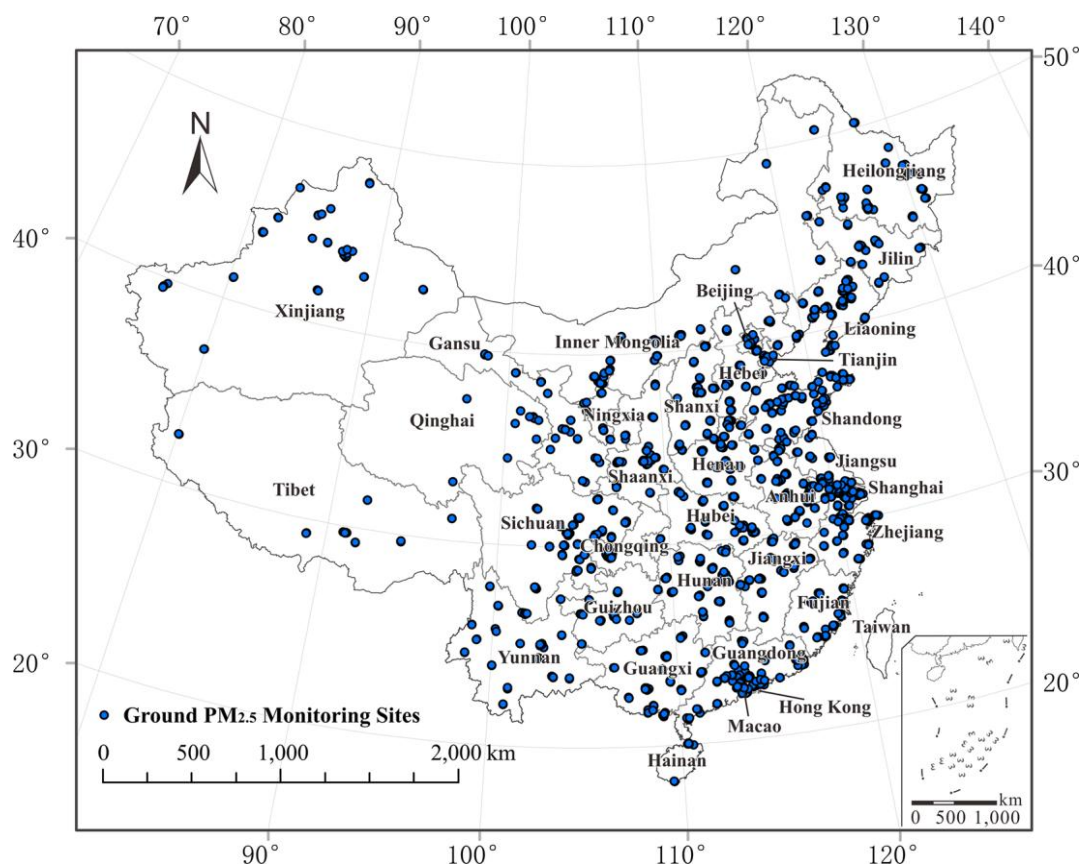

**Fig. 1.** Spatial distribution of PM$_{2.5}$ monitoring sites in China, as of the end of 2014.

*2.2. MODIS AOD products*

The Moderate Resolution Imaging Spectroradiometer (MODIS) aboard the National Aeronautics and Space Administration (NASA)'s Earth Observing System (EOS) satellites, Terra and Aqua, can provide retrieval products of aerosol and cloud properties with nearly daily global coverage (Remer et al., 2005). Recently, a new version of MODIS aerosol products (Collection 6) is released at a higher spatial resolution (3 km at nadir). Both the newly released 3-km and the prior standard 10-km AOD products are retrieved using the dark target algorithm, whereas a single retrieval box of 6 × 6 pixels and 20 × 20 pixels is averaged, respectively. Moreover, the pixels outside the reflectivity range of the brightest 50% and darkest 20% at 0.66 $\mu m$ are discarded to reduce uncertainty for the 3-km AOD products (Levy et al., 2013; Livingston et al., 2014).

In our study, both MODIS Terra and Aqua 3-km AOD products, corresponding to the ground $PM_{2.5}$ measurements in space and time, are downloaded from Level 1 and Atmosphere Archive and Distribution System (LAADS) website (http://ladsweb.nascom.nasa.gov). The coverage of these two AOD products differs because of the different crossing times of the two sensors. Hence, linear regression analysis are conducted between those pixels where both AOD products are present for each day, and the regression coefficients are used to predict the missing Aqua AOD values from corresponding available Terra AOD, and vice versa (Hu et al., 2014b; Ma et al., 2014). Then the average of two AOD products are employed to estimate daily average $PM_{2.5}$ concentration.

*2.3. Meteorological data*

In our study, the NASA atmospheric reanalysis data called the second Modern-Era Retrospective analysis for Research and Applications (MERRA-2) (Molod et al., 2015; Rienecker et al., 2011) is used. It uses the Goddard Earth Observing System Model, Version 5 (GEOS-5) data assimilation system, which is able to use the newer microwave sounders and hyperspectral infrared radiance instruments, as well as other data types. The MERRA-2 meteorological data are available from 1980, with a spatial resolution of 0.625° longitude × 0.5° latitude. More details about the MERRA-2 data can be found at the website (http://gmao.gsfc.nasa.gov/GMAO_products/).

We extracted relative humidity (RH, %), air temperature at 2m height (TEMP, K), wind speed at 10 m above ground (WS, m/s), surface pressure (SP, Pa), and planetary boundary layer height (HPBL, m) between 10 am and 11 am local time (Terra satellite overpass time corresponds to 10:30 am local time), and 1 pm and 2 pm local time (Aqua satellite overpass time corresponds to 1:30 pm local time), respectively. Each meteorological parameter were averaged over the two period to supplement the predictors for the daily average estimation of $PM_{2.5}$ concentration.

*2.4. Data preprocessing and matching*

All data above are re-processed to be consistent temporally and spatially to form a complete dataset which serves as the foundational samples for model development. Firstly, the satellite AOD and meteorological reanalysis data are regridded to 0.3 degree. Secondly, all data are re-projected to the same projection coordinate system. Finally, ground $PM_{2.5}$ measurements are associated with the value of satellite AOD and meteorological data covering the station. The averaging over multiple pixels is expected to effectively reduce random errors, but the best size of a single window centered at a given $PM_{2.5}$ monitoring site still remains unclear for our analysis. Hence, three different window sizes of $1 \times 1$, $3 \times 3$, $5 \times 5$ pixels were applied here, respectively, and with consistence to previous studies (Wu et al., 2012; You et al., 2015b), the average scheme over a window size of $3 \times 3$ pixels reported a slight advantage. After data preprocessing and matching, a total of 77978 records from multi-source data, which spans almost 2 years and contains ground-level $PM_{2.5}$, satellite-based AOD, MERRA meteorological reanalysis data, were collected for model development.

**3. Methodology**

*3.1. Previous retrieval models*

Many different models have been developed to estimate $PM_{2.5}$ concentration at both regional and national scales, ranging from single-variable to multi-variable models, and including both linear and nonlinear models. In this study, the following widely used models were evaluated and compared.

*3.1.1. Corrected linear regression (CLR)*

The linear regression model was used in earlier studies to describe the AOD-PM relationship (Chu et al., 2003; Wang and Christopher, 2003). Later studies reported better performances after meteorological correction (Li et al., 2005; Wang et al., 2010). Hence, corrected linear regression model was used:

$$PM'_{2.5} = a + b \cdot \frac{AOD}{HPBL} \qquad (1)$$

$$PM'_{2.5} = PM_{2.5} \cdot \left(\frac{1}{1 - RH/100}\right) \qquad (2)$$

where AOD is the aerosol optical depth, and $PM'_{2.5}$ denotes the RH-corrected PM$_{2.5}$.

*3.1.2. Multiple linear regression (MLR)*

Through incorporating more meteorological parameters, MLR has been introduced into the prediction of PM$_{2.5}$ concentration (Benas et al., 2013; Gupta and Christopher, 2009b). Based on empirical statistics, it can be defined as:

$$PM_{2.5} = \beta_0 + \beta_1 \cdot AOD + \beta_2 \cdot TEMP + \beta_3 \cdot RH + \beta_4 \cdot WS + \beta_5 \cdot HPBL + \beta_6 \cdot SP \qquad (3)$$

where $\beta_0$ is the interception for PM$_{2.5}$ prediction, and $\beta_1 \sim \beta_6$ are regression coefficients for the predictor variables.

*3.1.3. Semi-empirical model (SEM)*

Based on related physical understanding and statistical theory, SEM was developed to describe the relationship between meteorological observations, AOD, and PM$_{2.5}$ (Emili et al., 2010; Liu et al., 2005; Tian and Chen, 2010; You et al., 2015a). It can be expressed as:

$$PM_{2.5} = e^{\beta_0 + \beta_2 \cdot TEMP + \beta_3 \cdot RH} \cdot AOD^{\beta_1} \cdot WS^{\beta_4} \cdot HPBL^{\beta_5} \tag{4}$$

*3.1.4. Geographically weighted regression (GWR)*

The GWR model was developed to account for the spatial heterogeneity of the AOD-$PM_{2.5}$ relationship (Hu et al., 2013; Ma et al., 2014; Song et al., 2014; You et al., 2015b). Unlike the previous models, GWR does not predict $PM_{2.5}$ concentration using globally constant parameters, but generates continuous parameters by local model fitting. It can be represented as Eq. (5):

$$PM_{2.5,i} = \beta_{0,i} + \beta_{1,i} \cdot AOD + \beta_{2,i} \cdot TEMP + \beta_{3,i} \cdot RH + \beta_{4,i} \cdot WS + \beta_{5,i} \cdot HPBL + \beta_{6,i} \cdot SP \tag{5}$$

where the meanings of the variables and coefficients are the same as Eq. (3), but based on local regression over monitoring station $i$, hence the coefficients vary in space. In our study, the adaptive bandwidths were used because of the uneven distribution of the $PM_{2.5}$ stations.

*3.1.5. Back-propagation neural network (BPNN)*

An ANN can be considered as a set of computer algorithms designed to simulate biological neural networks in terms of machine learning and pattern recognition. Hence, ANNs have the potential to extract trends in imprecise and complicated nonlinear data (Gupta and Christopher, 2009a). With more and more predictors, the estimation of $PM_{2.5}$ concentration has been gradually considered to be a multi-variable and nonlinear problem. Consequently, ANNs have been introduced into $PM_{2.5}$ estimation (Gupta and Christopher, 2009a; Wu et al., 2012; Yao and Lu, 2014). The most common training algorithm is back-propagation (BP), a BPNN model with three layers (input layer, hidden layer, and output layer) was constructed in our study. The input parameters were latitude, longitude, month, AOD, TEMP, RH, WS, HPBL, and SP. According to previous studies (Gardner and Dorling, 1998; Reich et al., 1999), the number of nodes in the hidden layer ranges from $2\sqrt{n} + \mu$ to $2n+1$, where $n$ and $\mu$ are the number of nodes in the input layer and output layer, respectively. Thus, the number of nodes in the hidden layer was varied from 7 to 19, and 18 nodes (which performed the best) were selected in this paper.

*3.2. Proposed generalized regression neural network (GRNN) model*

With extensive and comprehensive data, an ANN has the potential to describe the spatial and temporal variation of AOD-$PM_{2.5}$ relationship. Previous studies (Gupta and Christopher, 2009a; Ordieres et al., 2005) have indicated that ANNs can outperform the classic statistical models. However, the well-known BPNN has the disadvantages of slow convergence velocity and easily converging to local minimum (Wen et al., 2000; Yu, 1992). Hence, we introduce here another neural network named the generalized regression neural network (GRNN), which can overcome the shortcomings of BPNN (Cigizoglu and Alp, 2006; Kisi and Kerem Cigizoglu, 2007). GRNN is often used for function approximation, and it can be considered as a normalized radial basis function (RBF) network. Based on a standard statistical technique called kernel regression, GRNN can solve any function approximation problem if sufficient data are given. A common GRNN architecture has three layers of neurons: the input layer, the RBF hidden layer, and the special linear output layer. The input layer and RBF hidden layer are usually connected by a density function such as the Gaussian density function. The output of the hidden layer is not directly connected to the output layer by a linear function, but is firstly connected by a transition of a dot function, reflecting the specialty of the output layer. Further theoretical details about GRNN can be found in previous studies (Specht, 1991; Specht, 1993).

In our study, the input signals are latitude, longitude, month, AOD, TEMP, RH, WS, HPBL,

and SP, and the output parameter is PM$_{2.5}$ concentration. The main function of the GRNN model is to estimate a nonlinear regression surface of PM$_{2.5}$ from these input signals. The input data are sent to the neural network, the output PM$_{2.5}$ concentrations can be calculated. They are compared with the in situ PM$_{2.5}$ data, and an error is estimated. The error is sent back to the GRNN model to adjust the weights to generate a more appropriate surface of PM$_{2.5}$ concentration. This process is therefore to find optimal value of weights to establish the nonlinear relationship between PM$_{2.5}$ and independent predictors. The GRNN model here predicts PM$_{2.5}$ concentration using AOD and meteorological parameters, which are primarily and supplementary predictors, respectively. In particular, allowing for the temporal and spatial variation of AOD-PM$_{2.5}$ relationship, the latitude, longitude, and month are also input to better estimate PM$_{2.5}$ concentration. Unlike BPNN, the number of nodes in the hidden layer of GRNN was obtained from training without artificial intervention; hence, GRNN was trained with few parameters set in advance. The schematics of GRNN used to estimate PM$_{2.5}$ concentration are presented in Fig. 2.

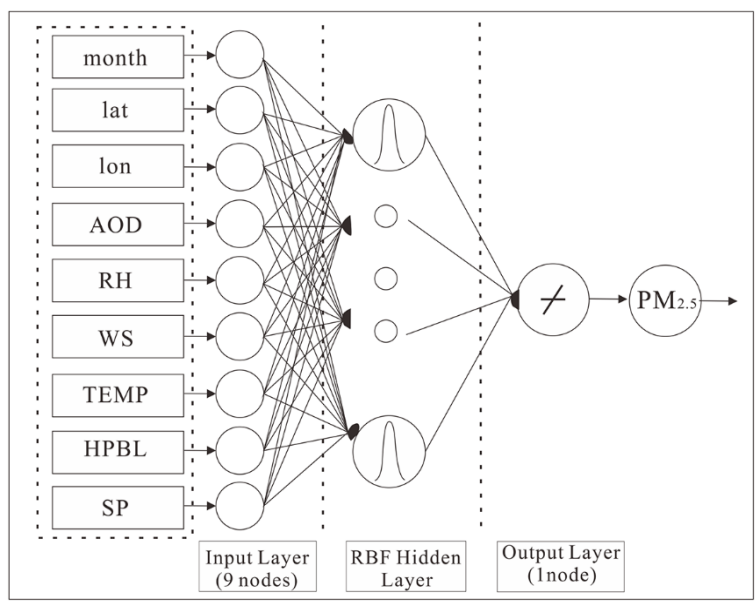

**Fig. 2.** Schematic of GRNN used to estimate PM$_{2.5}$ concentration in China.

*3.3. Model validation*

To validate the performance of the above models, the correlation coefficient (R) and root-mean-square error (RMSE) between the observed $PM_{2.5}$ and estimated $PM_{2.5}$ were adopted. Furthermore, a 10-fold cross-validation method (Rodriguez et al., 2010) was applied to test the model overfitting and predictive power. The dataset was averagely divided into 10 folds randomly. Nine folds of the dataset were used for model fitting, and one fold was predicted in each round of the cross-validation. This step was repeated 10 times until every fold was tested. Finally, R and RMSE values were calculated for the model fitting and cross-validation results, respectively, to evaluate the model performance.

*3.4. Mapping strategy for the mean $PM_{2.5}$ distribution*

The 3-km MODIS AOD products, retrieved by the dark target algorithm, share a generic feature with other standard AOD products, which is the absence of data due to clouds or high surface reflectance. Due to the absence of AOD data, the temporally continuous $PM_{2.5}$ data cannot be retrieved at the same location. When we map the temporal mean distribution of $PM_{2.5}$ concentration, a direct average can only reflect the level of $PM_{2.5}$ pollution on certain days. To address this issue, a pixel-based merging strategy is proposed, referring to the related studies of spatiotemporal fusion (Shen et al., 2013; Wu et al., 2013; Wu et al., 2015). Firstly, a spatial $PM_{2.5}$ map on every day can be interpolated by ground-station-level measurements. Due to the sparse distribution of the stations, this map is relatively coarse in space, but still keeps the temporal trend. Thus, the variation of the interpolated $PM_{2.5}$ remains equal to that of the satellite-derived $PM_{2.5}$ at the same location during the same period, and can be represented as Eq. (6),

$$L_{i,p} - L_{i,m} = F_{i,p} - F_{i,m} + \varepsilon_{i,m,p} \tag{6}$$

where $L_{i,m}$ denotes the PM$_{2.5}$ at pixel $i$ on day $m$ interpolated by ground-station-level measurements using the inverse distance weighting (IDW) approach, and $F_{i,m}$ is the satellite-based estimation of PM$_{2.5}$ at pixel $i$ on day $m$, as are $L_{i,p}$ and $F_{i,p}$. $\varepsilon_{i,m,p}$ is the random error, which will be reduced or eliminated by temporally averaging. For pixel $i$, assuming that $F_{i,p}$ is missing and needs to be estimated, then $m$ is the closest day with valid PM$_{2.5}$ data at the same location to day $p$. Thus, the missing satellite-derived PM$_{2.5}$ data can be reconstructed, and we can map the mean distribution of PM$_{2.5}$ concentration in China.

Additionally, some previous studies (van Donkelaar et al., 2012; Zheng et al., 2016) used annual/seasonal mean surface PM$_{2.5}$ measurements to correct this potential sampling biases. They calculate correction factors for each monitoring station/grid, then the factors are extrapolated to the entire study region. By applying the correction factors for each pixel/grid to predict annual/seasonal average PM$_{2.5}$, the bias-corrected PM$_{2.5}$ concentration can be obtained. Hence, our mapping strategy is compared with this bias-correction method (we denote it as "BCM").

**4. Results and discussion**

*4.1. Assessment of the various models*

*4.1.1. Performance of the models*

Table 2 shows the performance of various models. In model fitting, R values range from 0.485 to 0.895, and RMSEs from 16.51 to 31.64 $\mu g / m^3$. In the cross-validation results, a similar trend appears. Using simple linear regression, the CLR model performs the worst, with R and RMSE values of 0.488 and 31.51 $\mu g / m^3$ for cross-validation, respectively.

There is a large improvement (0.488 to 0.531 for R) from the CLR model to the MLR model, which considers more meteorological factors. Through introducing some physical prior knowledge, the SEM model has an advantage in $PM_{2.5}$ estimation, with an R value of 0.552 and an RMSE of 30.80 $\mu g/m^3$, respectively. Unlike the above models, the GWR model incorporates spatial information into the AOD-$PM_{2.5}$ relationship, and shows a large improvement over the SEM model, with R increasing by 0.058 and RMSE decreasing by 2.12 $\mu g/m^3$, respectively. As one of the intelligent algorithms, the BPNN model has the capacity to better represent the AOD-$PM_{2.5}$ relationship, with R and RMSE values of 0.693 and 25.96 $\mu g/m^3$, respectively. Compared with the results of BPNN, R increases by 0.123 (from 0.693 to 0.816), and RMSE decreases by 5.03 $\mu g/m^3$ (from 25.96 to 20.93 $\mu g/m^3$) for cross-validation of the GRNN model. These findings suggest that the proposed GRNN model performs the best, followed by BPNN and GWR, and then SEM and MLR, whereas the simple CLR model obtains the worst performance at national scale.

Table 2. Performance of the various models.

|      | Model fitting (N=70180) | | Cross-validation (N=77978) | |
| --- | --- | --- | --- | --- |
|      | R | RMSE ($\mu g/m^3$) | R | RMSE ($\mu g/m^3$) |
| CLR  | 0.485 | 31.64 | 0.488 | 31.51 |
| MLR  | 0.530 | 30.52 | 0.531 | 30.52 |
| SEM  | 0.551 | 30.78 | 0.552 | 30.80 |
| GWR  | 0.624 | 28.23 | 0.610 | 28.68 |
| BPNN | 0.699 | 25.90 | 0.693 | 25.96 |
| GRNN | 0.895 | 16.51 | 0.816 | 20.93 |

Furthermore, it should be noted that the conventional models (CLR, MLR, and SEM) have all obtained reasonable results at regional scale in China (Li et al., 2005; Song et al., 2014; Wang et al., 2010), but do not perform well at national scale, with R values of 0.488~0.552 and RMSEs of 30.80~31.51 $\mu g/m^3$ for cross-validation. The results indicate that the conventional models cannot adequately represent the association between PM$_{2.5}$ and independent variables at national scale. However, with R values ranging from 0.610 to 0.816 and RMSEs from 20.93 to 28.68 $\mu g/m^3$, the more advanced models (GWR, BPNN, and GRNN) have a great advantage in the estimation of PM$_{2.5}$ concentration.

Allowing for the superiority of the GRNN model, we further evaluated and analyzed its performance. Fig. 3 shows the scatter plots for GRNN model fitting and cross-validation. The R and RMSE values for model fitting are 0.895 and 16.51 $\mu g/m^3$, respectively. From model fitting to cross-validation, R decreases by 0.079 and RMSE increases by 4.42 $\mu g/m^3$. The results demonstrate that the proposed model results in a slight overfitting (Hu et al., 2014a; Ma et al., 2014). However, with the highest R and lowest RMSE values for both model fitting and cross-validation, the proposed GRNN model outperforms the other models. Hence, despite a slight overfitting, the GRNN model is more effective for the estimation of PM$_{2.5}$ concentration in China.

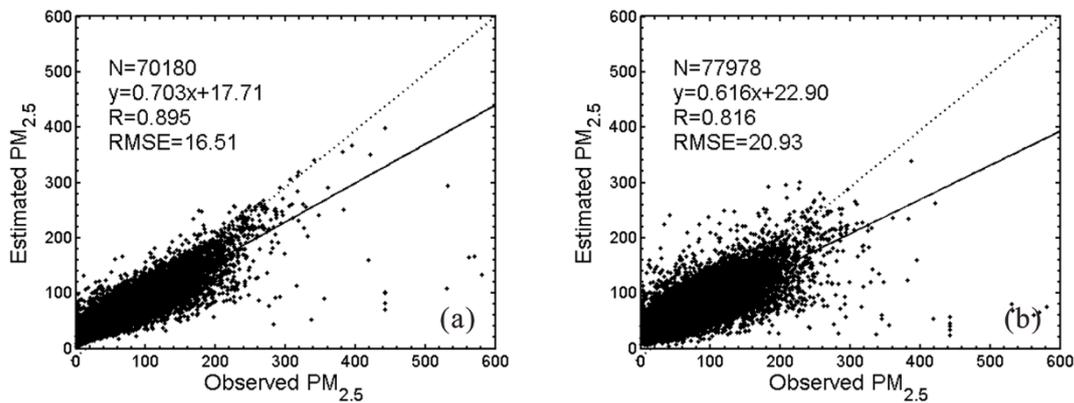

**Fig. 3.** Results of GRNN (a) model fitting and (b) cross-validation. PM$_{2.5}$ unit: $\mu g/m^3$. The dashed line is the 1:1 line as the preference.

To further analyze the spatial performance of the GRNN model, the R and RMSE values between the observed and estimated PM$_{2.5}$ over the stations were calculated and are presented in Fig. 4. The correlation coefficients at 727 out of 828 stations are greater than 0.8, and 84.90% of the total stations report a low RMSE of less than 20 $\mu g/m^3$. Spatially, the higher R values are clustered in Eastern China, suggesting the accurate estimation of PM$_{2.5}$ concentration in this area. In contrast, the lower R values are found in Northwest China, which is probably caused by the sparse distribution of the ground stations in this area. Moreover, a higher RMSE cluster appears in the Beijing-Tianjin-Hebei (BTH) region and its surroundings. However, it should be noted that the level of PM$_{2.5}$ concentration in the BTH region is relatively high (Lin et al., 2015; Ma et al., 2014).

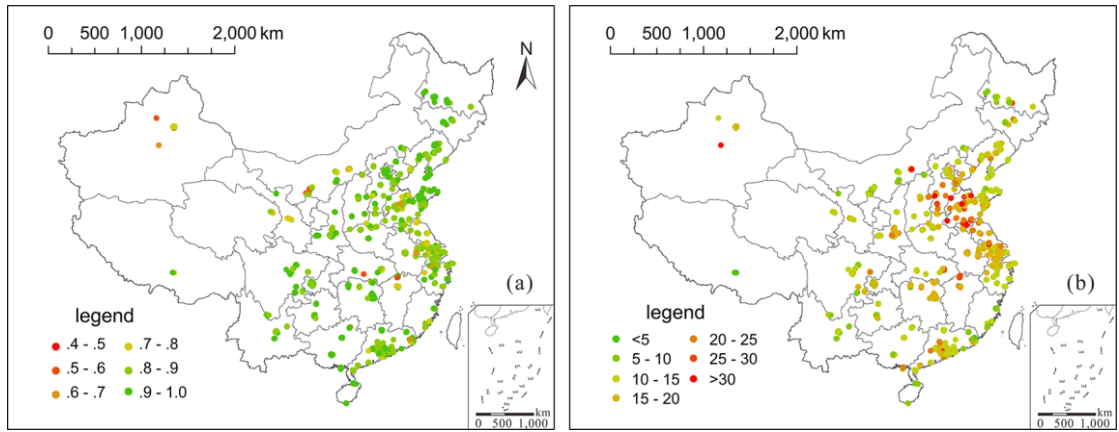

**Fig. 4.** Spatial distribution of (a) R and (b) RMSE ($\mu g/m^3$) between observed and estimated PM$_{2.5}$ over the stations.

*4.1.2. Seasonal variation of model performance*

Previous studies have shown that the models can perform differently as a function of the seasons (Gupta and Christopher, 2009a; Lin et al., 2015). Therefore, all the models were respectively established in each season to discuss and compare the seasonal variation of model performance. In our study, the seasons were defined as spring (March–May), summer (June–August), autumn (September–November), and winter (December–February), for which the numbers of data records were 21573, 23244, 26281, and 6880, respectively. Fig. 5 shows the seasonal variation of model performance for cross-validation.

As shown in Fig. 5, the GRNN model performs the best in every season (R= 0.822, 0.817, 0.828, and 0.837 for cross-validation in spring, summer, autumn, and winter, respectively), followed by BPNN and GWR, and then MLR and SEM, and the simple CLR model obtains the poorest performance (the R values of the four seasons are 0.351, 0.587, 0.569, and 0.502, respectively). The results demonstrate that by taking more meteorological parameters into consideration, the relatively advanced models can significantly improve the accuracy of $PM_{2.5}$ estimation. Furthermore, for each season, the GWR model performs a little worse than BPNN, but better than the conventional models, indicating that incorporating spatial information into the statistical model can better describe the AOD-$PM_{2.5}$ relationship. Among four seasons, all models have achieved the poorest performance in spring, probably caused by the influence of the enhanced contribution of dust particle (Zhang and Cao, 2015).

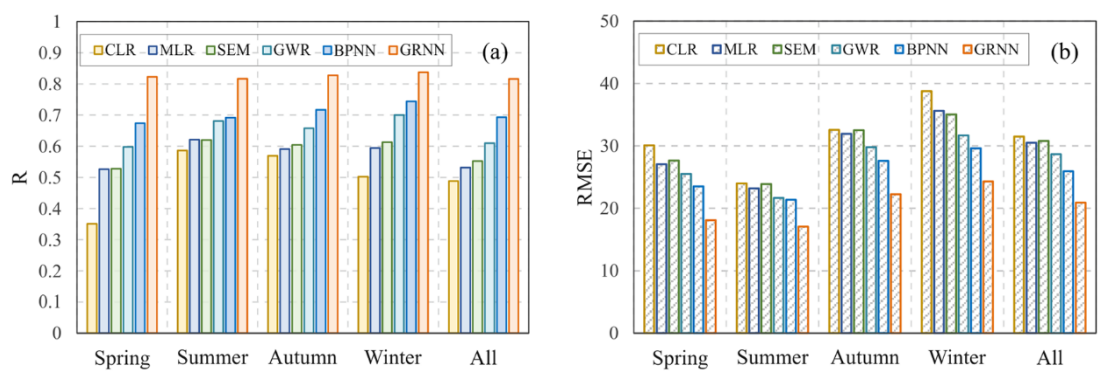

**Fig. 5.** Seasonal variation of (a) R and (b) RMSE ($\mu g/m^3$) between observed and estimated $PM_{2.5}$.

On the other hand, all the models generally perform better at seasonal scale (except for spring) than yearly scale, reflecting the influence of the seasons on the AOD-$PM_{2.5}$ relationship. It should be noted that the conventional models (CLR, MLR, and SEM) reports a more significant improvement of R value from yearly scale to seasonal scale, indicating that they seem to be more suitable for seasonal observation than yearly observation.

*4.1.3. Geographical variation of model performance*

Geographical location is considered to be one of the factors which has an influence on the AOD-PM$_{2.5}$ relationship (Hoff and Christopher, 2009; Ma et al., 2014). Clearly, China has a wide geographical range, and hence the models may perform differently in different regions. To explore the geographical variation of model performance, all the models were respectively established in every 4 °×4 ° grid box, as in the study of Gupta and Christopher (2009b). All the data measured at the stations falling in each grid box were collected. However, some grid boxes containing only a few (<4) stations were eliminated. Fig. 6 shows the geographical variation of R values between observed and estimated PM$_{2.5}$ for cross-validation, and the summary of the R and RMSE statistics over all the grid boxes is presented in Table 3.

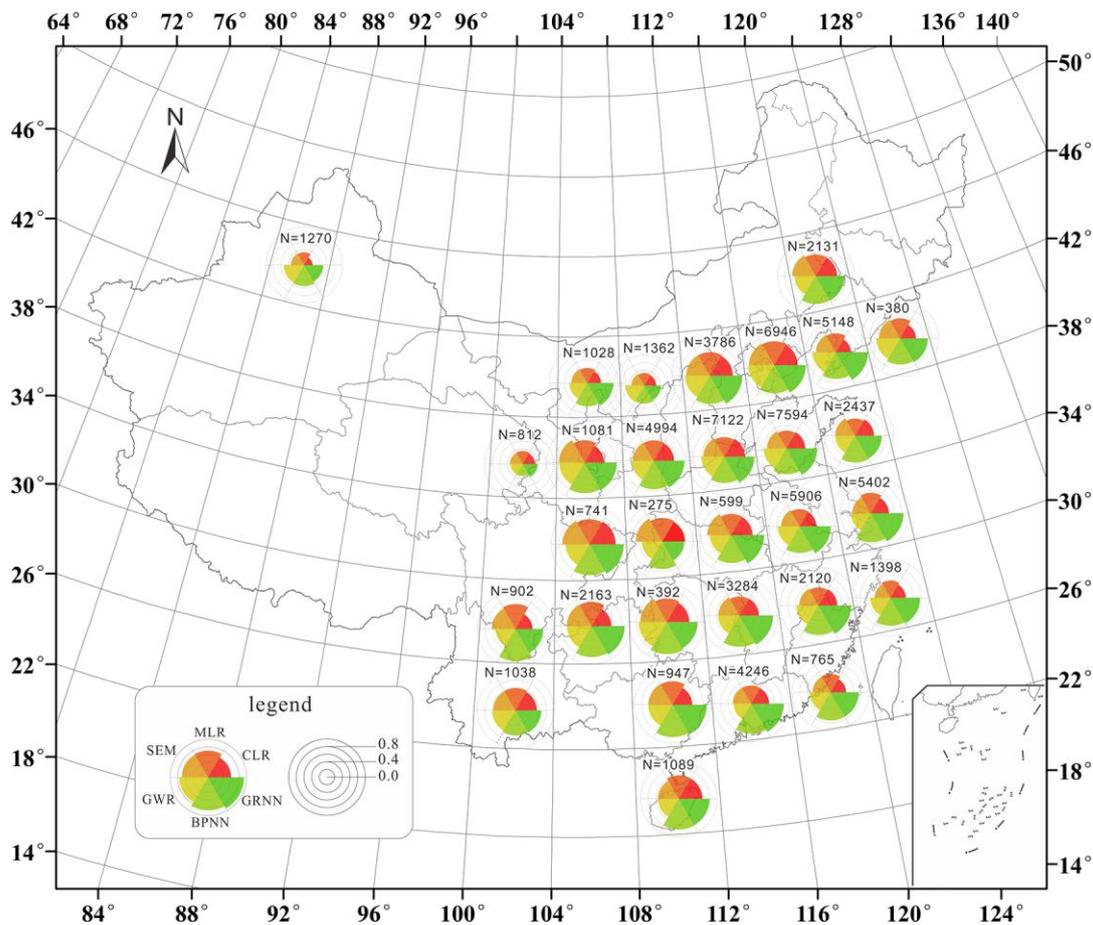

**Fig. 6.** Geographical variation of R values between observed and estimated PM$_{2.5}$ in each 4 °×4 ° grid box.

Compared with the performance of the various models at national scale, a similar trend can be investigated in most of the grid boxes. That is, the GRNN model performs the best, followed by BPNN and GWR, and CLR gives the poorest performance. However, the GWR model does not report great advantages over the conventional models, and the CLR model obtains almost the same performance as MLR. Additionally, some spatial differences are found. The GRNN model tends to obtain poorer results over the grid boxes in West China. This may be caused by the relatively sparse distribution of the $PM_{2.5}$ monitoring stations in this area. The models, especially the neural network models (BPNN and GRNN), achieve the highest accuracies and stability of performance over those grid boxes which locate between longitudes 112° and 120°, which contain more data records. These findings indicate that with more comprehensive data, the neural network models can perform better accordingly.

Table 3. Summary of the R and RMSE statistics over all the grid boxes.

|  | R | | | | | | RMSE ($\mu g/m^3$) | | | | | |
|---|---|---|---|---|---|---|---|---|---|---|---|---|
|  | CLR | MLR | SEM | GWR | BPNN | GRNN | CLR | MLR | SEM | GWR | BPNN | GRNN |
| Minimum | 0.219 | 0.328 | 0.336 | 0.320 | 0.317 | 0.386 | 14.87 | 14.58 | 14.75 | 13.49 | 11.64 | 12.30 |
| Maximum | 0.681 | 0.657 | 0.710 | 0.739 | 0.848 | 0.887 | 37.55 | 36.73 | 37.51 | 35.77 | 31.31 | 35.34 |
| Mean | 0.482 | 0.523 | 0.546 | 0.568 | 0.719 | 0.754 | 27.41 | 26.54 | 26.53 | 25.46 | 21.37 | 20.26 |
| Sd | 0.103 | 0.095 | 0.105 | 0.079 | 0.113 | 0.130 | 5.34 | 5.61 | 5.43 | 5.49 | 4.72 | 4.76 |

As Table 3 shows, the R value of the GRNN model ranges from 0.386 to 0.887, with a mean value of 0.754. There is a large decrease (0.816 to 0.754) from the national R value to the mean R value of the geographical grid boxes for cross-validation. However, an opposite trend appears for BPNN. On the other hand, the GRNN model obtains the highest standard deviation (Sd = 0.130) of R, meaning the biggest spatial variation of model performance. Meanwhile, the GWR model reports the smallest standard deviation (Sd = 0.079) of R, indicating that it is less sensitive to spatial location.

*4.2. Mapping the mean distribution of PM$_{2.5}$ concentration*

In Fig. 7, the yearly mean distributions of PM$_{2.5}$ concentration in China are mapped. The interpolated map presents a relatively coarse distribution of ground-level PM$_{2.5}$ concentrations, although it does not have much detailed spatial information, but is usually considered as a reference. The direct averaging of the satellite-derived PM$_{2.5}$ concentration does not share a similar spatial distribution with the interpolated map, with the most obvious difference being that Guangxi province has almost the same level of PM$_{2.5}$ as the BTH region. The reason for this is that the satellites can only detect the PM$_{2.5}$ pollution on certain days on which the AOD data are available. The results based on the proposed merging strategy share a similar spatial pattern with the interpolated map, but with many more details. Overall, the results suggest that the proposed pixel-based merging strategy is more effective for mapping the distribution of PM$_{2.5}$ concentration in China.

Based on the results obtained by the proposed merging strategy, a further analysis was undertaken. It can be seen that the spatial distribution of PM$_{2.5}$ concentration in 2014 is very similar to that in 2013. In most regions of China, the variation of PM$_{2.5}$ concentration falls in a range of −5 to 5 $\mu g/m^3$ from 2013 to 2014, indicating that China was generally suffering from the same level of PM$_{2.5}$ pollution from 2013 to 2014. However, some spatial differences should be noted. There is a relatively large decrease (10~20 $\mu g/m^3$) in the junction of Hebei, Shandong, and Henan provinces, but a contrasting trend can be seen in the junction of Hubei and Chongqing.

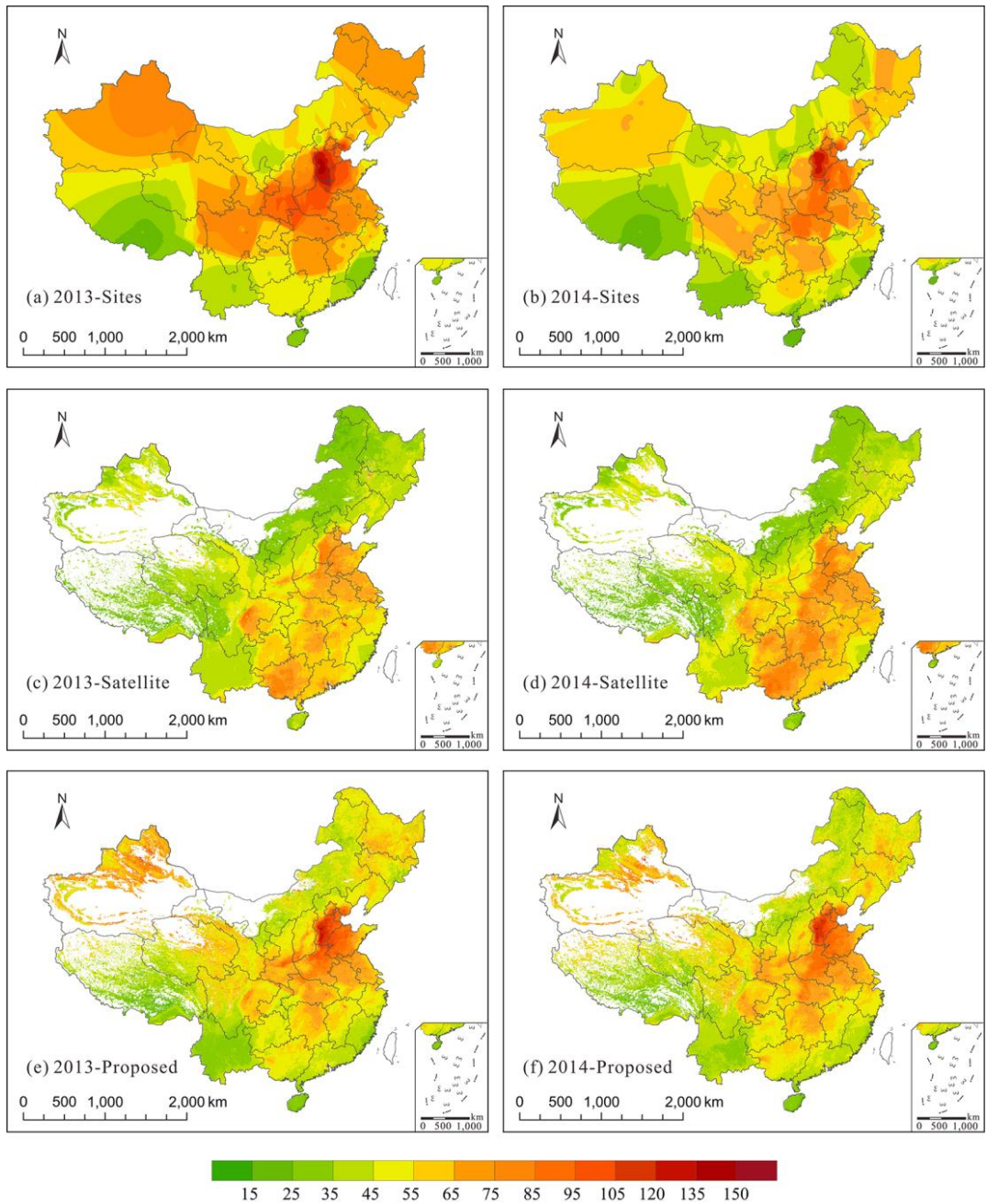

**Fig. 7.** Annual mean distribution of PM$_{2.5}$ concentration ($\mu g / m^3$) in China, (top) the results interpolated by ground site measurements, (middle) the direct averaging of satellite-derived PM$_{2.5}$ concentration, and (bottom) the mean distribution based on the proposed pixel-based merging strategy.

Spatially, the PM$_{2.5}$ pollution in West China is not as serious as that in Eastern China, which is accordance with the distribution of economic development and urbanization. Moreover, a strong north-to-south decreasing gradient is found, which agrees with the findings of previous studies (Lin et al., 2015). It should be noted that inner China generally suffers from a heavier

PM$_{2.5}$ pollution level than the southeastern coast; for instance, Central China (Hunan, Hubei, and Henan provinces) has a higher level of PM$_{2.5}$ concentration than Guangdong and Fujian provinces. In particular, a highly polluted region is located in the North China Plain, with a yearly average PM$_{2.5}$ concentration of about 85~120 $\mu g/m^3$. Previous studies (Quan et al., 2011; Tao et al., 2012) showed that rapid industrialization and urbanization have led to serious PM$_{2.5}$ pollution in this area. The cleanest regions are Hainan province, part of Yunnan province, and Tibet, with yearly mean PM$_{2.5}$ concentrations of less than 35 $\mu g/m^3$.

The seasonal mean distribution of PM$_{2.5}$ concentration in China was also mapped using the proposed pixel-based merging strategy. As presented in Fig. 8, 2013 and 2014 show similar seasonal trends. Winter is the most polluted season, whereas summer is the cleanest. According to previous studies (Han et al., 2010; Yu et al., 2011), this may be caused by winter heating in North China, Northeast China, and Northwest China.

The distribution of PM$_{2.5}$ concentration is qualitatively similar to the station measurements. To make a further validation of the results, the R and RMSE values between the yearly and seasonal mean mapped PM$_{2.5}$ and in situ PM$_{2.5}$ were calculated, respectively. As Table 4 shows, the proposed merging strategy obtains better results than BCM at yearly and seasonal scale, with the R values all greater than 0.90, suggesting that the mapped PM$_{2.5}$ distribution quantitatively agrees quite well with the station measurements. On the other side, the BCM method performs worst in winter, with the reason that many stations, on which the AOD data is missing during the whole winter, cannot be used for bias-correction. In fact, the BCM method has a systematic drawback being that many station measurements cannot be used for both model development and bias-correction due to the AOD absence during the whole study period, whereas the proposed method can make the best use of the station measurements.

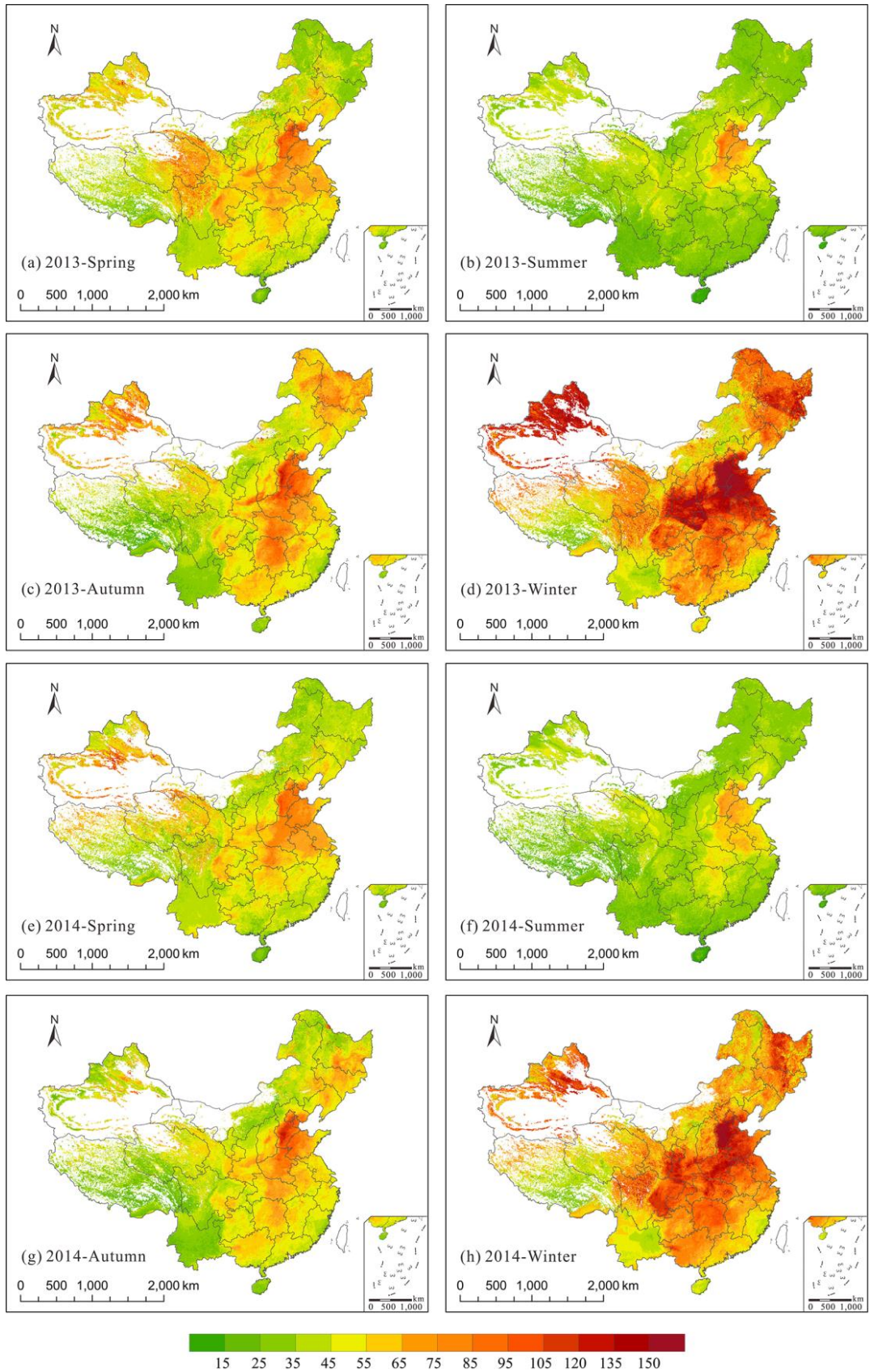

**Fig. 8.** Seasonal mean distribution of PM$_{2.5}$ concentration ($\mu g / m^3$) in China.

Table 4. R and RMSE ($\mu g / m^3$) values between yearly and seasonal mean mapped PM$_{2.5}$ and in situ PM$_{2.5}$.

| method | | 2013 | | | | | 2014 | | | | |
|---|---|---|---|---|---|---|---|---|---|---|---|
| | | All | Spring | Summer | Autumn | Winter | All | Spring | Summer | Autumn | Winter |
| Proposed | R | 0.964 | 0.930 | 0.955 | 0.941 | 0.957 | 0.946 | 0.930 | 0.930 | 0.924 | 0.951 |
| | RMSE | 6.73 | 7.55 | 6.44 | 9.07 | 13.85 | 6.93 | 7.31 | 6.30 | 8.94 | 10.05 |
| BCM | R | 0.935 | 0.897 | 0.926 | 0.870 | 0.838 | 0.917 | 0.914 | 0.882 | 0.837 | 0.814 |
| | RMSE | 8.56 | 8.72 | 8.07 | 12.72 | 25.62 | 8.18 | 8.04 | 7.76 | 13.08 | 21.80 |

*4.3. Discussion: comparison with previous studies*

To date, there are two strategies that have been used for point-surface fusion for the estimation of ground-level PM$_{2.5}$ concentration. One strategy is that the models are established based on all the data records collected from the whole study period (Gupta and Christopher, 2009a; Lin et al., 2015; Wu et al., 2012; Yao and Lu, 2014); the other strategy focuses on a daily basis (Ma et al., 2014; Ma et al., 2016; Song et al., 2014; Xie et al., 2015). Hence, our study can be classified as the former strategy. This strategy can predict the historical PM$_{2.5}$ concentrations which cannot be provided by ground station measurements, whereas the latter strategy has an advantage in the real-time monitoring of PM$_{2.5}$ pollution. To further validate our study, it was compared with one of the former studies (Lin et al., 2015) which focused on a national scale. To some degree, the model performance of the two strategies cannot be intercompared because of the huge differences, but we still made some attempts to qualitatively compare our results with those of the latter studies (Ma et al., 2014; Ma et al., 2016).

A good correlation (R=0.9) between annual mean observed and estimated PM$_{2.5}$ in 2013 was reported in a previous study (Lin et al., 2015). Our results show a slight advantage in PM$_{2.5}$ estimation, with R values between the GRNN-estimated PM$_{2.5}$ and the corresponding observed PM$_{2.5}$ of 0.929 during the same period. We also focused on the spatial pattern of PM$_{2.5}$ concentration in China. Our results are very similar to theirs, but are slightly lower along the Bo Hai Coast and higher in Shaanxi province. On the other hand, we also attempted to compare our results with those based on a daily basis and a grid technique (Ma et al., 2014; Ma et al., 2016). The R and RMSE values of these results are 0.80/0.89 and 32.98/27.42 $\mu g/m^3$, whereas we report 0.816 and 20.93 $\mu g/m^3$, respectively. There is a decrease in R value from their latter study to ours. However it should be noted that their study and ours have many significant differences in data and methods; for example, the AOD gap filling was undertaken in their study. Hence, the variation of R/RMSE cannot be the whole story. Compared with their former study, our results share a similar spatial distribution. However, a slightly higher level of PM$_{2.5}$ concentration in the BTH region and a lower level in the Sichuan Basin are reported. Furthermore, our spatial pattern is also very like their results of the 10-year (2004–2013) PM$_{2.5}$ mean distribution. The North China Plain has the highest level of PM$_{2.5}$ concentration, and a gradual decrease appears from the north to the south.

## 5. Conclusions

To sum up, our study has several benefits and advantages. Firstly, we have introduced the new generalized regression neural network (GRNN) model to better describe the AOD-PM$_{2.5}$ relationship. Secondly, the performance of various widely used models was evaluated and compared at national scale. Finally, a pixel-based merging strategy was proposed to effectively map the yearly and seasonal mean distribution of PM$_{2.5}$ concentration in China.

With cross-validated R values of 0.488~0.552 and RMSEs of 30.80~31.51 $\mu g/m^3$, the conventional models did not report good results at national scale, although they did obtain some reasonable results under certain conditions in previous studies. In contrast the more advanced models achieved better performances in PM$_{2.5}$ estimation, with R values of 0.610~0.816 and RMSEs of 20.93~28.68 $\mu g/m^3$. The proposed GRNN model obtained the best results, with the highest R (0.816) value and lowest RMSE (20.93 $\mu g/m^3$) among all the models. The R values between the yearly/seasonal mean mapped PM$_{2.5}$ and observed PM$_{2.5}$ were all greater than 0.90, indicating that the mapped PM$_{2.5}$ distribution agrees quite well with the station measurements. This study therefore has the capacity to provide reasonable information for the spatiotemporal analysis of PM$_{2.5}$ variation.

In future studies, we will focus on three aspects. Firstly, statistical methods will be introduced into filling the missing AOD data (Li et al., 2015; Shen et al., 2015; Zeng et al., 2013), because a wider coverage of satellite-based AOD could provide more comprehensive information for PM$_{2.5}$ estimation. Secondly, we will take more variables into consideration; for example, land use and population. More parameters associated with PM$_{2.5}$ pollution could lead to an improvement in PM$_{2.5}$ estimation accuracy. Finally, a long-term analysis of PM$_{2.5}$ pollution in China will be made to facilitate epidemiological studies about the impact of air pollution on public health, using the estimated PM$_{2.5}$ concentration at a 3-km resolution.

**Acknowledgments**

We would like to acknowledge the data providers of the Chinese National Environmental Monitoring Center (CNEMC), Goddard Space Flight Center Distributed Active Archive Center (GSFC DAAC) and US National Aeronautics and Space Administration (NASA) Data Center, respectively.